\documentstyle[preprint,aps]{revtex}
\tightenlines
\begin{document}
\draft
\title{Structure of Vacuum Condensates}
\author{L.S. Kisslinger and T. Meissner}
\address{Department of Physics, Carnegie Mellon University, 
Pittsburgh, PA 15213, U.S.A.}
\title{Structure of Vacuum Condensates}

\date{\today} 

\maketitle
 
\begin{abstract}
It is essential to know the space-time structure of the nonlocal vacuum 
condensates for application to medium energy processes. Using the 
Dyson-Schwinger formalism in the rainbow approximation for the quark
propagator, we study the nonlocal quark condensate and model forms for 
the nonperturbative gluon propagator constrained by fits to local condensates
and deep inelastic scattering with nucleon targets. 
\end{abstract}


\section{Introduction}
It has long been known that chiral symmetry breaking requires a
nonperturbative quark propagator with non vanishing vacuum matrix elements 
of normal ordered products of quark fields \cite{gor}, called quark 
condensates.  These condensates would vanish in a perturbative vacuum but do
not vanish in the QCD vacuum, and are of central importance both for the
structure of hadronic matter and for the study of the early universe chiral
phase transition. A systematic treatment of hadronic masses can be carried
out \cite{svz,rry,nar}
using operator product expansions in terms of vacuum matrix
elements of local operators, the vacuum condensates, whose phenomenological
values have been confirmed in lattice gauge calculations \cite{lg}.
As was discussed in early work on the magnetic dipole moments of
nucleons \cite{by}, for application to form factors and transition matrix
elements in the low to medium momentum transfer region the operator product
expansion cannot be used, since long distance properties of nonlocal
operators must be treated.    
 
One approach to this problem of treating bilocal operators has been the 
use of nonlocal condensates, which have been introduced to
represent the  bilocal vacuum matrix elements needed for the pion wave
function \cite{mr} and pion form factor \cite{br} for low to medium momentum
transfer. In this method one does not carry out an O.P.E. for the vacuum
matrix elements of the bilocal operators,
but introduces new phenomenological functions needed to characterize the
space-time structure of the nonlocal condensates. Both the forms of these
functions and the parameters are found by fits to experiment as well as
considerations of analyticity.  E.g., in a study of parton distribution
functions \cite{jk} the space-time scale of a nonlocal condensate was
determined by a fit of a monopole form in space-time to experimental data.
On the other hand, in a recent use of nonlocal condensates to determine
the values of vacuum susceptibilities \cite{jk1}, which characterize 
the nonperturbative quark propagation in an external field \cite{is},
it was found that the monopole form did not have suitable analytic 
properties, and a space-time dipole form was used to fit the low-x parton 
data.  Although a satisfactory fit to the phenomenological pion 
susceptibility \cite{hhk1} was found, it is a good example of the importance 
of determining the structure of the nonlocal condensates for application to
transition matrix elements over a wide range of momentum transfer.

It is the goal of the present work to study the form of the nonlocal quark
condensate using the the QCD Dyson-Schwinger [D-S]
equations \cite{iz,rw}.  Using the bare gluon-quark vertex,
defined as the rainbow approximation, the nonperturbative (dressed) quark
propagator is determined self consistently with a model for the 
nonperturbative (dressed) gluon propagator $D_{\mu\nu} ^{ab} (q)$.
A comprehensive review of this type of model is given in Ref. \cite{tandy}
It has been shown that in the rainbow approximation the value
of the quark condensate \cite{cahill,fm1} and the mixed quark condensate
\cite{t} can be obtained with suitable choices of the gluon propagator,
which also provides constraints for the present work.
The gluon condensate within this approach has been studied in 
Ref. \cite{tf}. 
Other studies use the D-S formalism with different
approximations \cite{gh} to attempt to determine the nonperturbative quark
condensates.  

For hadronic properties such as the elastic and transition form factors one
needs the information equivalent to the bound-state Bethe-Salpeter [B-S]
equation. It has also been shown that the rainbow D-S model is consistent
with nontopological chiral quark models \cite{cahill} and low-momentum
transfer meson form factors \cite{brt}.
For a treatment of form factors over an extended 
range of momentum transfer light-cone B-S studies of the pion form factor
have shown \cite{kw} that nonperturbative as well as perturbative two-quark
propagators are needed, and that in a QCD treatment four-quark matrix elements
are also required.  We plan a study of nonlocal nonperturbative four-quark
matrix elements within this approach in the near future.
 
In the present work we use the rainbow Dyson-Schwinger equation to
investigate the forms of the nonlocal quark condensate as well as
the gluon propagators.  Using the same phenomenological gluon propagators
as were used in previous studies of the local condensates \cite{cahill,fm1,t}
we find that the dipole form with the parameter close to the one
found from fits to the sea-quark distribution \cite{jk,jk1} can be obtained.
 
\section{Nonlocal Quark Condensate From An Effective Interaction In The
Dyson-Schwinger Approach}
 
The quark propagator is defined by
\begin{eqnarray}
 S_q(x) & = & <0|T[q(x)\bar{q}(0)]|0>, 
\label{eq-qp}
\end{eqnarray}
where $q(x)$ is the quark field and T the time-ordering operator. 
For the physical vacuum the quark propagator 
$S_q(x)$ has a perturbative and a nonperturbative part.
In the case of vanishing current quark masses ($m_0 =0$) one can write
\begin{eqnarray}
 S_q(x)  & = & S_q^{PT}(x) + S_q^{NP}(x),
 \nonumber\\
  S_q^{PT}(x) & = & \frac{1}{2\pi^2} \frac{\gamma\cdot x}{x^4}
 \nonumber\\
 S_q^{NP}(x) & = &  (-) \frac{1}{12} \left ( <:\bar{q} (x) q(0): > + 
x_\mu <:\bar{q} (x) \gamma^\mu q(0): >  \right )
\label{eq-qprop}
\end{eqnarray}
It should be stressed that
normal-ordered products, and therefore $S_q^{NP}$,
do not vanish in the nonperturbative vacuum.  
For short distances, the
O.P.E. for the scalar part of $S_q^{NP}(x)$ gives
\begin{equation}
<:\bar{q} (x) q(0): >    =  
<:\bar{q} (0) q(0): > -
 \frac{x^2}{4} 
<0|:\bar q (0) \sigma\cdot G(0) q(0) :|0> + \dots ,
\label{eq-ope}
\end{equation}
in which the local operators of the expansion are the quark condensate, 
the mixed condensate, and so forth.
 
In Ref. \cite{jk1} it is shown that with a choice of nonlocal 
condensate
\begin{equation}
<0|:\bar{q} (x) q(0):|0>  =  {\rm g}(x^2) <0|:\bar q(0) q(0):|0>,
\label{eq-nlc1}
\end{equation}
with
\begin{eqnarray}
 {\rm g}(x^2) & = & \frac{1}{(1+\kappa^2x^2/8)^2} 
\nonumber \\
        & = & \int_{0}^{\infty} d\alpha {\rm f}(\alpha)e^{-x^2\alpha/4},
\nonumber \\
 {\rm f}(\alpha) & = & \frac{4}{\kappa^4} \alpha e^{-2\alpha/\kappa^2},
\label{eq-f}
\end{eqnarray}
and $\kappa^2  = 0.15 \dots 0.20 {\mbox{GeV}}^2$, 
one can fit the low-$x$ quark distributions and
also the pion susceptibility. 

In the Dyson-Schwinger formalism $S_q^{NP}$ is related to the quark 
self-energy, $\Sigma$, by
\begin{equation}
 S_q(p)^{-1}  =  i\gamma \cdot p + \Sigma(p).
\label{eq-sig}
\end{equation}
Using the bare quark gluon vertex (the rainbow approximation),
$\Gamma_\nu ^b(q,p) = \gamma_\nu \frac{\lambda_c ^b}{2}$, $\Sigma(p)$
satisfies the rainbow Dyson-Schwinger equation \cite{rw}:
\begin{equation}
 \Sigma(p)  =  \int \frac{d^4q}{(2\pi)^4}  g_s ^2 D_{\mu\nu} ^{ab} (p-q) 
\gamma_\mu \frac{\lambda_c ^a}{2} S_q(q)\gamma_\nu \frac{\lambda_c ^b}{2}
\label{sd}
\end{equation}
with $D_{\mu\nu} ^{ab} (q)$ the gluon propagator,
$\lambda_c ^a$ the color $SU(3)$ matrix. 
In Euclidean space one can write
\begin{equation}
S_q(p)^{-1}  =  i\gamma \cdot p A(p^2) + B(p^2) 
\label{eq-ab}
\end{equation}

The choice of the Landau gauge for the gluon propagator
\begin{equation} 
D^{ab}_{\mu\nu}(q) = \delta^{ab}
\left ( \delta_{\mu\nu} - \frac{q_\mu q_\nu}{q^2} \right ) D(q^2)
\label{landau}
\end{equation}
leads to the set of coupled integral equations
\begin{eqnarray}
[A(p^2) - 1]p^2 & = & \frac{4}{3} g_s ^2 
\int \frac{d^4q}{(2\pi)^4} D(p-q)
\frac{A(q^2) }{q^2A^2(q^2) + B(q^2) ^2}
\left [ p\cdot q + 2 \,\frac{(p\cdot q-q^2)(p^2-p\cdot q)}{(p-q)^2} \right ] 
\nonumber\\
B(p^2) & = & 4 g_s ^2 \int \frac{d^4q}{(2\pi)^4} D(p-q)
\frac{B(q^2)}{q^2A^2(q^2) + B(q^2) ^2}   \, .
\label{eq-sd}
\end{eqnarray}
 
Mesonic bound states can be studied within this framework by solving the
ladder Bethe Salpeter equations for the corresponding ${\bar q} q$ bound 
states. 
Various mesonic properties have been studied in Refs.\cite{cahill}.
In Ref.\cite{fm1} a detailed investigation of the low energy sector was
performed by deriving the general form of the effective chiral action for the
SU(3) Goldstone bosons and determining $f_\pi$ and most of the chiral low
energy coefficients $L_i$, which, in turn, determines the physics of the
$\pi$, $K$ and $\eta$ mesons at low energies \cite{chipth}.

Because the form of the gluon propagator $D(s)$
in the IR region is unknown, we must use model forms as input.
Our model ansatz is
\begin{equation}
{g_s}^2 D(s) \,= \, 
3 \pi^2  \frac{\chi^2}{\Delta^2} e^{-\frac{s}{\Delta}} \, ,
%\frac{4\pi \alpha_s (s)}{s} \, = \, 
\label{gluon}
\end{equation}
which determines the quark-quark interaction through a strength parameter
$\chi$ and a range parameter $\Delta$.
Its form is inspired by the $\delta$ function ansatz of Ref. \cite{mn},
which it approaches for $\Delta \to 0$.
 
The nonlocal quark condensate 
$<:\bar{q} (x) q(0) :>$ is then
given by the scalar part of the Fourier transformed
inverse quark propagator:
\begin{eqnarray}
<:\bar{q} (x) q(0) :>  &=& 
(- 4 N_c) \int
\frac{d^4 p}{(2\pi)^4} \frac{B(p^2)}{p^2 A^2(p^2) + B^2 (p^2) } e^{ipx}
\nonumber \\
&=& (-) \frac{12}{16 \pi^2} \int_0 ^\infty 
ds s \frac{B(s)}{s A^2(s) + B^2 (s)}
\left [ 2 \frac{J_1 (\sqrt{s x^2})}{\sqrt{s x^2} } \right ]
\label{fourier}
\end{eqnarray}
At $x=0$ the expression
for the local condensate $<:\bar{q} q : >$ is recovered:
\begin{equation}
<:\bar{q} q : > = 
 (-) \frac{12}{16 \pi^2} \int_0 ^\infty
ds s \frac{B(s)}{s A^2(s) + B^2 (s)}
\label{local}
\end{equation}
The nonlocality $g(x^2)$ can be obtained immediately by 
dividing (\ref{fourier}) through (\ref{local}).
 
Because the quark-quark interaction defined by (\ref{gluon})
has a finite range in momentum space the momentum integrals in
(\ref{fourier}) and (\ref{local}) are finite.
Our analysis ignores effects from hard gluonic radiative corrections to the
condensates which are connected to a possible change of the renormalization
scale $\mu$ at which the condensates are defined.
Those effects are of minor importance for our study of 
nonperturbative effects in the low and medium energy regions.
It should be stressed in this context that our
interaction is not renormalizable 
because we are using the bare quark gluon vertex.
Therefore, instead of our condensates
depending logarithmically on the renormalization scale $\mu$ \cite{hw},
the scale at which a condensate is defined in our approach is 
a typical hadronic scale, which is implicitly
determined by the model gluon propagator $g_s ^2 D(s)$ and the solutions
of the D-S equations (\ref{sd}). 
The situation is very similar to the determination of vacuum condensates
in the instanton liquid model \cite{pw,dem} where the scale is set 
by the inverse instanton size.
 
In order to check the sensitivity of our
results on the model gluon 2 point function (\ref{gluon})
we try various sets of parameters $\chi$ and $\Delta$
and investigate the $x^2$ dependence of the 
function $g(x^2)$ for these forms.
We solve the set of integral equations (\ref{eq-sd}) self consistently
for a given model form for ${g_s}^2 D(q^2)$
obtaining the quark propagator functions 
$A(p^2)$ and $B(p^2)$,
which, in turn will allow us to calculate $g(x^2)$ from (\ref{fourier}).     
The result can then be compared to the dipole fit of Ref.\cite{jk1}
with $\kappa^2 = 0.15 \dots 0.20 {\mbox{GeV}}^2$.
The parameter sets we are using are
\begin{eqnarray}
\mbox{Set 1:}    \; \Delta &=& 2.0 * 10^{-3} \mbox{GeV}^2 \; ; \;  \chi = 1.40
\mbox{GeV}\nonumber \\
\mbox{Set 2:}    \; \Delta &=& 1.0 * 10^{-2} \mbox{GeV}^2 \; ; \;  \chi = 1.56
\mbox{GeV}\nonumber \\
\mbox{Set 3:}    \; \Delta &=& 2.0 * 10^{-2} \mbox{GeV}^2 \; ; \;  \chi = 1.58
\mbox{GeV}\, .
\nonumber \\ 
\label{sets}
\end{eqnarray}
These parameters have been chosen so that they
reproduce the correct value for the pion decay constant in the chiral
limit $f_\pi = 88 \mbox{MeV}$.
Moreover the values 
of the chiral low energy coefficients $L_i$ \cite{chipth} are compatible 
with the phenomenological values in both cases.
Following Ref.\cite{fm1} one finds :
$L_1 = 0.86$, $L_3 = -4.53$, $L_5 = 0.78$, $L_8 = 0.84$ ($*10^{-3}$) 
for Set 1;
$L_1 = 0.84$, $L_3 = -4.48$, $L_5 = 0.88$, $L_8 = 0.84$ ($*10^{-3}$)
for Set 2 and
$L_1 = 0.83$, $L_3 = -4.42$, $L_5 = 0.92$, $L_8 = 0.78$ ($*10^{-3}$)
for Set 3. 
 
Many of the works in Refs.\cite{tandy,cahill,fm1,t} have used 
a model ansatz for the gluon propagator
\begin{equation} 
D^{ab}_{\mu\nu}(q) =
\delta^{ab}\delta_{\mu\nu}
D(q^2) .
\label{feynmangauge}
\end{equation}
which is often referred to as the Feynman-like gauge.
It is however not identical to the Feynman gauge QCD,
in which the dressed gluon propagator would have different longitudinal
and transverse components. Therefore the ansatz (\ref{feynmangauge})  
should be regarded merely as a model form for the gluon 2 point function.
For our purpose it is, however, interesting to ask if and how
the nonlocal quark condensate
depends on choosing either Landau gauge (\ref{landau}) or 
the Feynman-like gauge (\ref{feynmangauge}).
Therefore we perform the calculation for another parameter set:  
\begin{equation}
\mbox{Set 4:}    \; \Delta = 2.0 * 10^{-3} \mbox{GeV}^2 \; ; \;  \chi = 1.23
\mbox{GeV} \, ,
\label{set3}
\end{equation}
while using (\ref{feynmangauge}) instead of (\ref{landau})
for the gluon propagator.
Set 4 has the same range parameter $\Delta$ than Set 1.
The strength parameter $\chi$ is slightly smaller in order to 
obtain the correct value of $f_\pi = 88 \mbox{MeV}$.
The chiral low energy coefficients are:  
$L_1 = 0.85$, $L_3 = -4.46$, $L_5 = 0.82$, $L_8 = 0.93$ ($*10^{-3}$),
values rather close to those obtained with Set 1.

\section{Results and Discussion}
 
Fig.\ref{fig1} shows the results for $g(x^2)$ 
for the four parameter sets and compares with the
dipole fit of Ref.\cite{jk1} with $\kappa^2 = 0.20 \mbox{GeV}^2$ (solid line).
The result of Ref.\cite{jk1} is best reached for 
a gluon propagator with a small range parameter $\Delta = 0.002 \mbox{GeV}^2$ 
in the infrared.
Larger values for the width parameter $\Delta$ lead to stronger deviations
from the form of Ref.\cite{jk1}.
 
By comparing the curves for Set 1 and Set 4
we can demonstrate that the nonlocal condensate is very robust
with respect to using 
Landau gauge (\ref{landau}) or the Feynman-like gauge (\ref{feynmangauge}). 
The change between the two forms for the gluon 2 point function
can be easily made up by a slight readjustment
of the parameters of the IR model ansatz without any significant 
change of the final result.    
 
We conclude that the D-S formalism is a valuable tool for the study of
the nonlocal quark condensate, and expect that the B-S formalism will
also prove to be useful for the study of the nonlocal four-quark
condensates, which provide nonperturbative QCD effects for hadron
couplings and form factors. 

\acknowledgements
This work was supported in part by the National Science Foundation grant
PHY-9319641.

%\end{thebibliography}
 
\begin{figure}
\caption{The non local quark condensate 
$g(x) = < : {\bar{q}} (x) q(0): > / <: {\bar{q}} (0) q(0): > $
for the four sets of model gluon propagators mentioned in the text
compared with the dipole fit of 
Ref.{\protect\cite{jk1}}.}
\label{fig1}
\end{figure}
 

\begin{references}
%\begin{thebibliography}{99}
\bibitem{gor}
M. Gell-Mann, R.J. Oakes and B. Renner, Phys.Rev. {\bf 175}, 
2195 (1968).
\bibitem{svz} 
M.A. Shifman, A.I. Vainshtein and V.I. Zakharov, 
Nucl.Phys. B {\bf 147}, 385 (1979); 448 (1979).
\bibitem{rry} 
L. Reinders, H. Rubinstein and S. Yazaki, 
Phys. Rep. {\bf 127}, 1 (1985).
\bibitem{nar} S. Narison, 
{\it QCD Spectral Sum Rules}, World Scientific, Singapore, 1989, 
and references therein.
\bibitem{lg}
I. Montray and G. M\"unster, {\it Quantum Fields on a Lattice},
Cambridge University Press, 1994.
\bibitem{by} 
I.I. Balitsky and A.V. Yung, Phys.Lett. B {\bf 129}, 328 (1983).
\bibitem{mr} 
S.V. Mikhailov and A.V. Radyushkin, 
JETP Lett. {\bf 43}, 712 (1986); Phys.Rev. D {\bf 45}, 1754 (1992).
\bibitem{br} 
A.P. Bakulev and A.V. Radyushkin, Phys.Lett. B {\bf 271}, 223 (1991).
\bibitem{jk} 
H. Jung and L.S. Kisslinger, Nucl.Phys. A {\bf 586}, 682 (1995).
\bibitem{jk1} 
M.B. Johnson and L.S. Kisslinger, hep-ph/9706376,
submitted to Phys.Rev. D. 
\bibitem{is} 
B.L. Ioffe and A.V. Smilga, Nucl.Phys. B {\bf 232}, 109 (1984).
\bibitem{hhk1} 
E.M. Henley, W-Y.P. Hwang and L.S. Kisslinger, 
Phys.Lett. B {\bf 367}, 21 (1996).
\bibitem{iz}
C. Itzykson and J-B. Zuber, 
{\it Quantum Field Theory}, McGraw-Hill, 1985.
\bibitem{rw}
C.D. Roberts and A.G. Williams, Prog.Part.Nucl.Phys. {\bf 33} 477 (1994),
and references therein.
\bibitem{tandy} P. Tandy, Prog.Part.Nucl.Phys. 
{\bf 39}, (1997) to appear, and references therein. 
\bibitem{cahill}
R.T. Cahill and C.D. Roberts, Phys.Rev. D {\bf 32}, 2419 (1985);
C.D. Roberts, R.T. Cahill and J. Praschifka, Ann.Phys. (N.Y.) {\bf 188}, 
20 (1988); 
M.R. Frank, P.C. Tandy and G. Fai, Phys. Rev. C {\bf 43}, 2808 (1991); 
\bibitem{fm1}
M.R. Frank and T. Meissner,
Phys.Rev. C {\bf 53}, 2410 (1996).
\bibitem{t}
T. Meissner, Phys.Lett. B {\bf 405}, 8 (1997).  
\bibitem{tf}
H.-B. Tang and R.J. Furnstahl, hep-ph/9502326.
\bibitem{gh} Xin-Heng Guo and Tao Huang, Commun.Theor.Phys. {\bf 17}, 455
(1992).
\bibitem{brt} C.J. Burden, C.D. Roberts and M.J. Thomson, Phys.Lett. B
{\bf 371}, 163 (1996).   
\bibitem{kw} L.S. Kisslinger and S.W. Wang, Nucl. Phys B {\bf 399}, 63 (1993);
Phys. Rev. D {\bf 54}, 5890 (1996).
\bibitem{chipth}
J. Gasser and H. Leutwyler, Ann.Phys.(N.Y.) {\bf 158}, 142 (1983);
Nucl.Phys. B {\bf 250}, 465 (1985); for reviews c.f. 
G. Ecker, Prog.Part.Nucl.Phys. {\bf 35}, 1 (1995);
V. Bernard, N. Kaiser and U.G. Meissner, Int.J.Mod.Phys. E {\bf 4}, 
193 (1995).
\bibitem{mn}
H.J. Munczek and A.M. Nemirowsky, Phys.Rev. D {\bf 28}, 181 (1983). 
\bibitem{hw}
F.T. Hawes and A.G. Williams, Phys.Rev. D {\bf 51}, 3081 (1995). 
\bibitem{pw}
M.V. Polyakov and C. Weiss, Phys. Lett. B {\bf 387}, 841 (1996).
\bibitem{dem}
A.E. Dorokhov, S.V. Esaibegian and S.V. Mikhailov,
Phys.Rev. D {\bf 56}, 4062 (1997).    
\end{references}
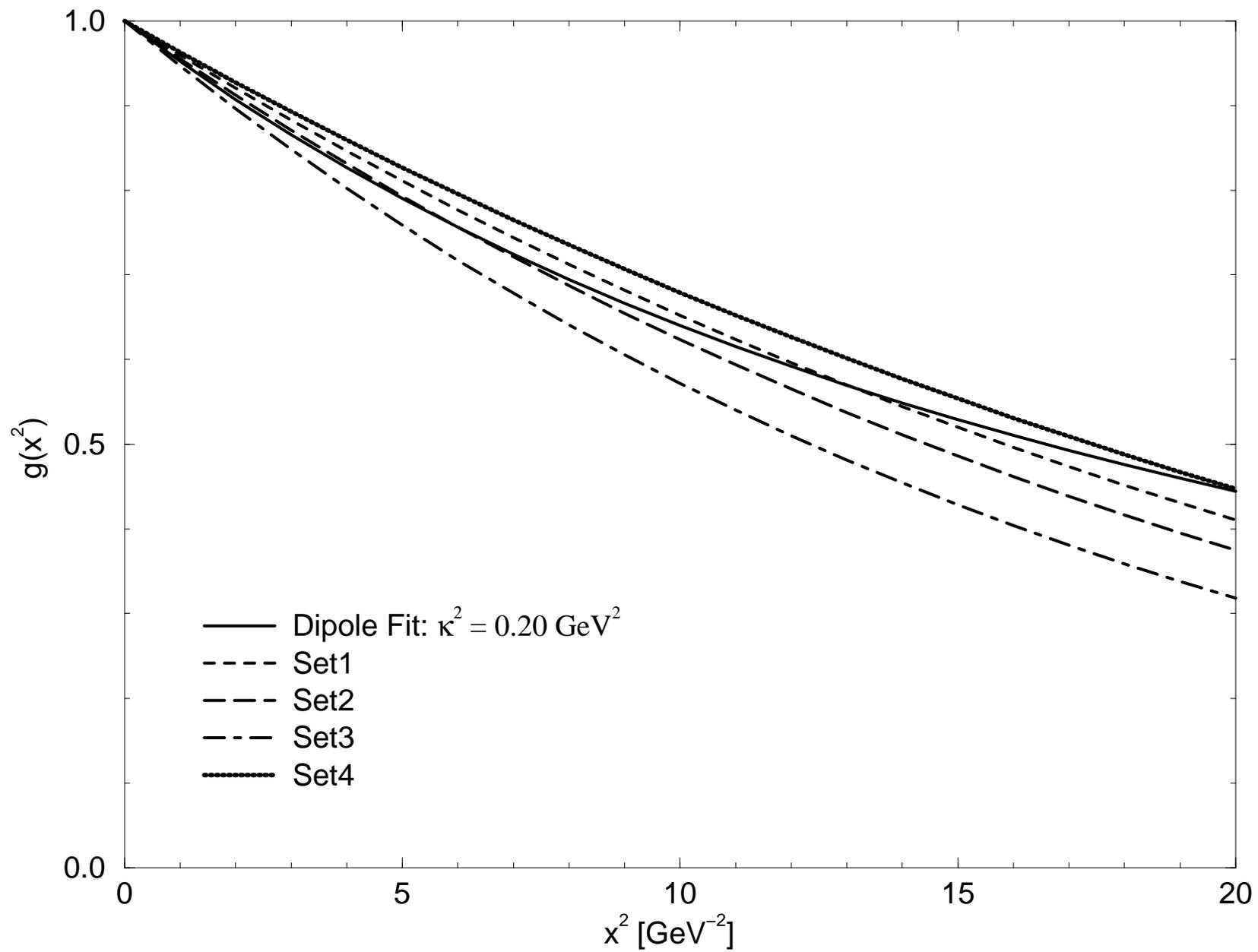
\end{document}